\documentclass[sigconf,nonacm=true,balance=true]{acmart}

\AtBeginDocument{%
  \providecommand\BibTeX{{%
    \normalfont B\kern-0.5em{\scshape i\kern-0.25em b}\kern-0.8em\TeX}}}

\setcopyright{none}

\usepackage{graphicx,subcaption}
\usepackage[autostyle]{csquotes}
\usepackage{listings}
\usepackage{color}
\definecolor{light-gray}{gray}{0.95}
\definecolor{codegray}{rgb}{0.5,0.5,0.5}
\usepackage{enumitem}
\usepackage{hyperref}
\hypersetup{
pdftitle={A Modular and Extensible Hardware Platform Prototype for Dynamic Data Physicalisationh},
pdfsubject={Modular and Extensible Hardware Platform Prototypes},
pdfauthor={Xuyao Zhang, Milan Ilić and Beat Signer},
pdfkeywords={Dynamic Data Physicalisation, Tangible Interaction, Plug-and-Play Architecture, Modular Hardware Platform, Rapid Prototyping}
}

\newcommand{\code}[1]{\texttt{#1}}

\lstdefinestyle{mystyle}{
  backgroundcolor=\color{backcolour}, commentstyle=\color{codegreen},
  keywordstyle=\color{magenta},
  numberstyle=\tiny\color{codegray},
  stringstyle=\color{codepurple},
  basicstyle=\ttfamily\footnotesize,
  breakatwhitespace=false,         
  breaklines=true,                 
  captionpos=b,                    
  keepspaces=true,                 
  numbers=left,                    
  numbersep=5pt,                  
  showspaces=false,                
  showstringspaces=false,
  showtabs=false,                  
  tabsize=2
}

\AtBeginDocument{%
  \providecommand\BibTeX{{%
    Bib\TeX}}}

\begin{document}

\title[A Modular and Extensible Hardware Platform Prototype for Dynamic Data Physicalisation]{A Modular and Extensible Hardware Platform Prototype\\ for Dynamic Data Physicalisation}

\author{Xuyao Zhang}
\orcid{0000-0002-5597-6217}
\affiliation{%
  \institution{WISE Lab}
  \department{Vrije Universiteit Brussel}
  \city{Brussels}
  \postcode{1050}
  \country{Belgium}
}
\email{xuyzhang@vub.be}

\author{Milan Ili\'c}
\affiliation{%
  \institution{WISE Lab}
  \department{Vrije Universiteit Brussel}
  \city{Brussels}
  \postcode{1050}
  \country{Belgium}
}
\email{milan.pavle.ili@vub.be}

\author{Beat Signer}
\orcid{0000-0001-9916-0837}
\affiliation{%
  \institution{WISE Lab}
  \department{Vrije Universiteit Brussel}
  \city{Brussels}
  \postcode{1050}
  \country{Belgium}
}
\email{bsigner@vub.be}


\begin{abstract}
Dynamic data physicalisation is an emerging field of research, investigating the representation and exploration of data via multiple modalities, beyond traditional visual methods. Despite the development of various data physicalisation applications in recent years, the integration of diverse hardware components remains both time-consuming and costly. Further, there is a lack of solutions for rapid prototyping and experimentation with different dynamic data physicalisation alternatives. To address this problem, we propose a modular and extensible hardware platform for dynamic data physicalisation.  This platform introduces a communication architecture that ensures seamless plug-and-play functionality for modules representing different physical variables. We detail the implementation and technical evaluation of a preliminary prototype of our platform, demonstrating its potential to facilitate rapid prototyping and experimentation with various data physicalisation designs. The platform aims to support researchers and developers in the field by providing a versatile and efficient tool for the rapid prototyping and experimentation with different data physicalisation design alternatives.
\end{abstract}




\maketitle


\section{Introduction and Related Work}
Dynamic data physicalisation is an emerging field of research that brings digital data into the physical world, allowing people to leverage their perceptual abilities for data exploration and understanding~\cite{jansen2015opportunities}. Numerous data physicalisation solutions have been developed to enhance data comprehension~\cite{taher2015exploring,daniel2019cairnform,Laina2021} and experiments have demonstrated that interacting with data in a physical form not only increases engagement but also makes the experience more enjoyable. 

In many cases, data physicalisation shares the same goal as data and information visualisation, both of them transforming raw data into an easily comprehensible structure. However, implementing data physicalisations is inherently more complex than visualisations, as it requires more technical work, including 3D~printing, hardware setup and cabling. Further, if a researcher or developer wishes to alter the method of data representation---such as using temperature instead of colour to convey a value---they must rebuild everything from scratch, which is a time-consuming and costly process. 

To address this challenge, we propose a modular and extensible hardware platform that facilitates the rapid prototyping of dynamic data physicalisations. The platform has a plug-and-play architecture, allowing researchers and users to create their own data physicalisations by simply connecting various modules to a core component. This eliminates the need to manually integrate different hardware and software components or determine how to use physical variables~\cite{jansen2015opportunities} to represent data attributes. The presented platform forms part of a more general research effort on realising a dynamic data physicalisation framework and pipeline, where a dedicated grammar informs the automatic mapping and transformation of data into physical variables forming part of a specific dynamic data phyicalisation~\cite{signer2018}, enabling new forms of human-information interaction~\cite{signer2024,signer2019}. In the following, we discuss some related work in the domain of dynamic data physicalisation and plug-and-play architectures.

\subsection{Dynamic Data Physicalisation} \label{dynamicdataphys}
A dynamic data physicalisation solution might be updated and modified in response to new data input or user interactions. There are multiple ways to realise these dynamic updates. A possible way to make data physicalisation dynamic is to combinine fabricated objects with augmented reality~(AR), as seen in molecular biology applications~\cite{gillet2005tangible,gillet2004augmented}, where virtual 3D~representations are superimposed to 3D~printed molecular models. Users can view different molecular properties by easily changing the virtual information. However, the physical parts are fixed and only the digital overlay information can be adapted. 

A more advanced solution involves shape-changing interfaces, which enable objects to change in various ways, such as orientation, form or texture~\cite{rasmussen2012shape}. These interfaces have been widely used in dynamic data physicalisation applications. Examples include FEELEX~\cite{iwata2001project}, a spatially continuous surface consisting of motorised pins that provide haptic feedback when users touch an image. Lumen~\cite{lumen} is another early interactive display consisting of an array of movable illuminated rods. 

Relief~\cite{leithinger2010relief,leithinger2011direct} goes one step further by allowing free-hand gesture input and direct touch input. Later examples such as inFORM~\cite{follmer2013inform} and TRANSFORM~\cite{ishii2015transform} support new interactions, including object actuation and remote gestural control. A more recent application, TiltStacks~\cite{tiab2018tiltstacks}, supports not only z-dimensional actuation, but also tilt movement, creating a more versatile display surface. Apart from that, Taher~et~al.~\cite{taher2015exploring} developed EMERGE, a dynamic physical bar chart that displays information from a dataset using a $10 \times 10$ array of self-actuating rods with RGB~LEDs, enabling both shape and colour output. In addition, a projector is used to show supplemental information on the table surface. All these examples implement shape-changing features based on an array of actuated rods, which is complex and time-consuming to realise and difficult to adapt if changes are needed.

Changing object shapes can also be achieved by using special materials. For instance, Follmer~et~al.~\cite{follmer2012jamming} proposed a particle jamming system to control material stiffness and provide both soft and rigid material states. PneUI~\cite{yao2013pneui} is another shape-changing interface using pneumatically actuated soft composite materials. However, these approaches also require substantial construction efforts and are limited by current developments in material science.     
\subsection{Plug-and-Play Architecture}
A good example to explain the idea of plug-and-play architectures is SensorBricks~\cite{SensorBricks}, a Lego-like toolkit designed for non-expert users to interact with data by reconfiguring different input and output bricks. For instance, a user might select a temperature input brick that measures room temperature and use a digital output brick to show the temperature on a segmented display. SensorBricks provides a fun and accessible way to interact with data, but it is limited in the supported modalities, and the position of each brick is unknown. 

Bloctopus~\cite{blikstein2015bloctopus} is another example of a plug-and-play architecture. It is a modular electronic prototyping toolkit that provides a general USB hub to which multiple sensors and actuators can be attached. Each module connected to the hub might communicate with a computer or a microcontroller by passing messages through MIDI. Although Bloctopus has not been used for data physicalisation, it demonstrates how a plug-and-play architecture can lower the entry barrier for building tangible interactions.

Similarly, Microsoft developed the .NET Gadgeteer project~\cite{villar2012netgadgeteer} to facilitate the quick and easy design and construction of custom electronic devices and systems. It consists of a main board to which various modules can be connected via a physical wire. Hardware compatibility is ensured by supporting multiple communication protocols such as UART, I$^2$C or SPI. However, a downside of this design is that all the sockets are located on the main board, implying that a larger main board with more sockets is required when more connections are needed. 

Finally, SoftMod~\cite{softmod2020} is a modular plug-and-play electronics kit for prototyping electronic systems. Similar to SensorBricks, it uses the concept of input and output modules interacting with each other via an I$^2$C~communication bus. The master module assigns addresses to each slave module, either directly or via other slave modules. An additional interesting feature is the ability to track the topology or spatial placement of modules relative to each other.

\section{Dynamic Data Physicalisation Platform}
To address the issues mentioned in the previous section, we propose a modular dynamic data physicalisation platform with a plug-and-play architecture for rapid prototyping. 

\begin{figure}[htb]
    \includegraphics[width=7.3cm]{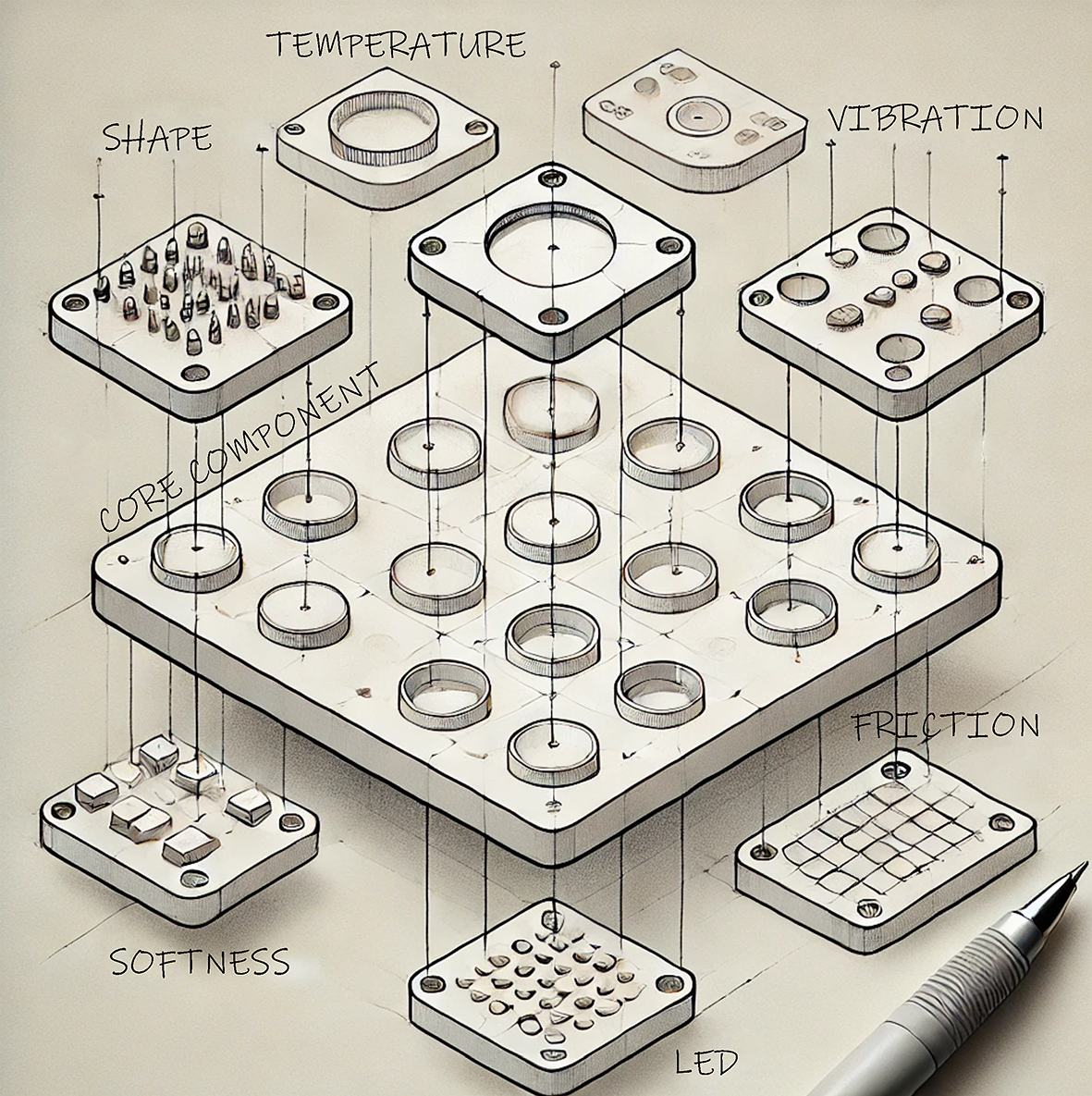}
    \caption{Concept sketch of the proposed data physicalisation platform (image created with the assistance of ChatGPT-4)}
    \label{fig:architecture}
\end{figure}

Figure~\ref{fig:architecture} shows a concept sketch of the proposed platform, which consists of a core component and multiple modules. Each module represents one or more physical variables, such as temperature or vibration, and uses specific hardware for the encoding. Modules can be connected to the core component via magnets and communicate with each other over a main bus. Once a module is connected, the core component is notified about what physical variable(s) the module represents, where it is connected and how it might be controlled. Subsequently, the core component can send commands to the module.

A key contribution of the proposed platform is the definition of the main bus for communication between modules and the core component. The bus should be scalable, capable of connecting additional modules, and able to integrate multiple data physicalisation platforms (i.e.~core components). Consequently, SPI and UART have not been further considered. Since both the modules and the core component should be able to initiate communication, buses using a master/slave protocol, such as USB and I$^2$C, were also discarded. The controller area network~(CAN)~\cite{hpl2002introduction}, commonly used in the vehicle industry~\cite{johansson2005vehicle}, meets our requirements as it is broadcast-based and message-oriented. By using a microprocessor with either a built-in or external CAN~controller and CAN transceiver, all modules and the core component can initiate connections and send messages. Adding more modules to the system simply involves connecting them to the main bus to enable communication. 

Despite the advantages of the CAN~bus, it cannot determine the spatial position of each module. To address this issue, we propose using Power Line Communication~(PLC)~\cite{hashim2023adaptation} to detect connected devices and assign a unique address to each connection point (slot). As the name suggests, Power Line Communication transmits data through power lines, a technology widely used in areas such as smart grids to reduce network construction costs. We leverage this benefit in our dynamic data physicalisation platform to provide both power and positional data to the connected modules, thereby saving on wiring costs. Most PLC modems support the I$^2$C protocol for automatic detection and we propose utilising these techniques.  

\begin{figure}[htb]
    \includegraphics[width=\columnwidth]{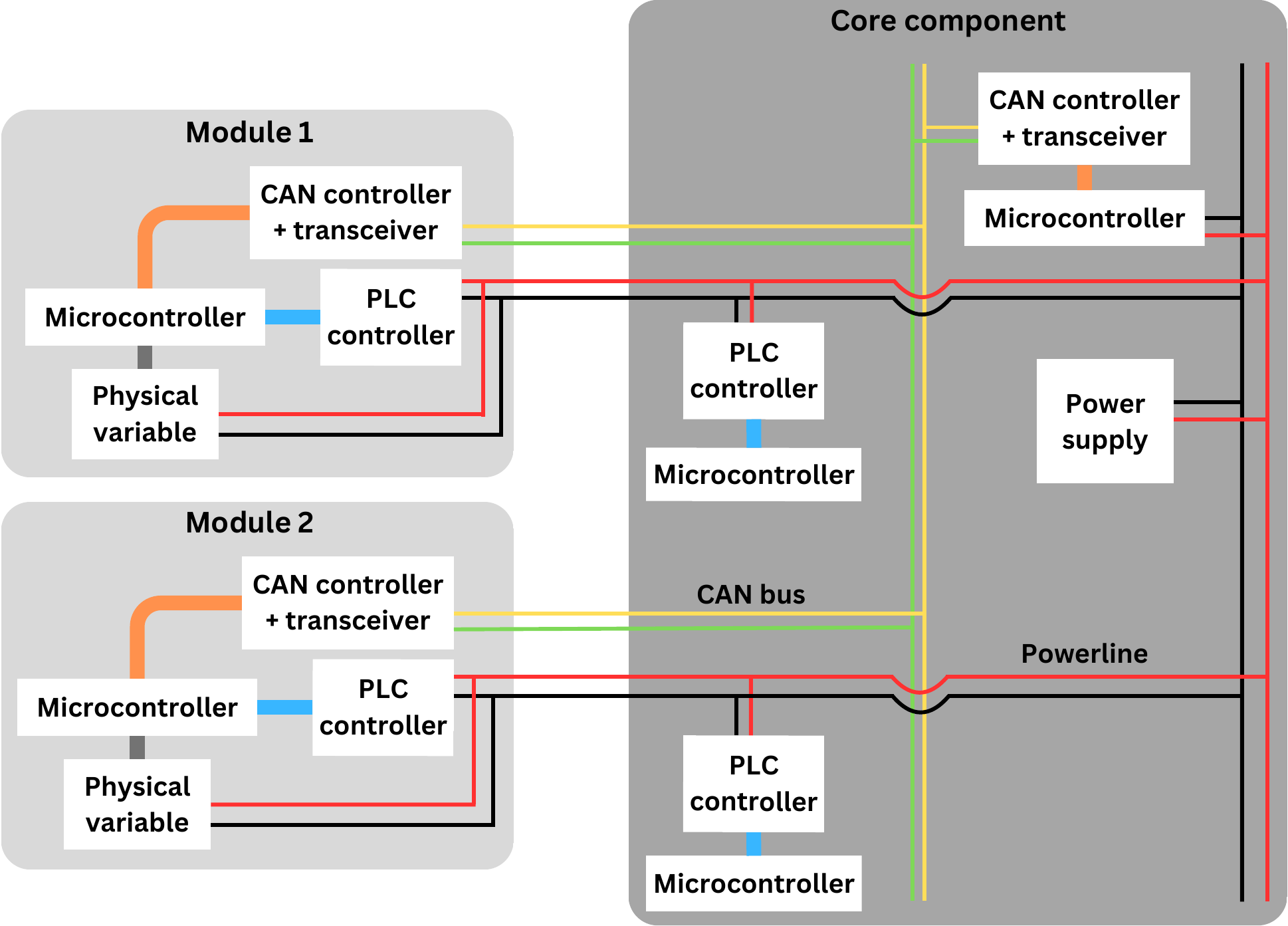}
    \caption{Communication architecture}
    \label{fig:solution}
\end{figure}

Figure~\ref{fig:solution} illustrates the communication architecture of the proposed solution. On the left, we present two example modules implementing one or more physical variables, such as temperature, vibration or shape. The main control unit of a module is a microcontroller, such as an Arduino Nano Every\footnote{https://store.arduino.cc/en-be/products/arduino-nano-every/}. This microcontroller connects to multiple hardware components~(e.g.~LED, stepper motor and vibration motor) to implement physical variables. The microcontroller is further connected to an external CAN controller and transceiver, as well as a PLC~controller. The former is used for CAN communication, while the latter deals with PLC communication. Therefore, a module's connector consists of four wires (green and yellow for CAN and black and red for power) connecting a module to the core component. The two PLC~wires also provide power to the microcontroller and any hardware (physical variables) it controls. The core component, shown on the right-hand side of Figure~\ref{fig:solution}, also contains a central microcontroller combined with a CAN controller and transceiver to parse messages sent over the CAN bus. Each slot of the core component is integrated with another microcontroller and a PLC~controller specifically providing a unique identifier (address) to a connected module. 

\begin{figure}[htb]
    \includegraphics[width=0.85\columnwidth]{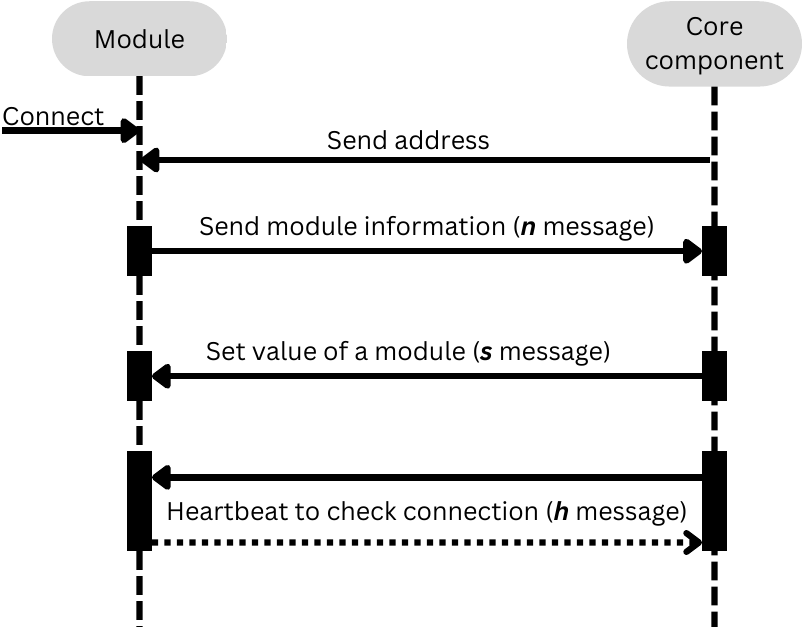}
    \caption{Communication of modules and core component}
    \label{fig:sequence}
\end{figure}

When a module is connected to a slot of the core component, it first receives a unique address via the PLC bus. This address is then used for communication over the CAN bus, as illustrated in Figure~\ref{fig:sequence}. The first message a module sends includes its description, detailing the physical variable(s) it encodes, the granularity (i.e.~number of different values it can encode) as well as any restrictions (e.g.~minimum and maximum values). Once this information is received, the core component can assign specific values to the module. Note that the core component also regularly sends some heartbeat messages to check whether a module is still connected.

We further propose a self-describing specification for the modules to provide information about the physical variables they encode and how the values of these variables can be adjusted (e.g.~minimum and maximum values or the number of individual values that can be encoded). The core component uses this information to set specific values in a module.

Based on the proposed architecture, we have implemented an initial prototype. We have chosen an Arduino Nano Every as the microcontroller for both the modules and core component due to its small form factor and low costs. Since the Arduino Nano Every lacks a built-in CAN controller, we combined it with an MCP2515 CAN bus module, which includes a CAN controller and transceiver. Our prototype focuses on CAN communication and does not yet implement the PLC part. Therefore, the two power wires are currrently only used for charging and we used a serial cable to connect a module's Arduinio GPIO pin to the one of the Arduino controlling the slot to transmit the unique slot address. 

\begin{figure}[htb]
     \centering
     \begin{minipage}{0.495\columnwidth}
         \centering
         \includegraphics[width=4.05cm]{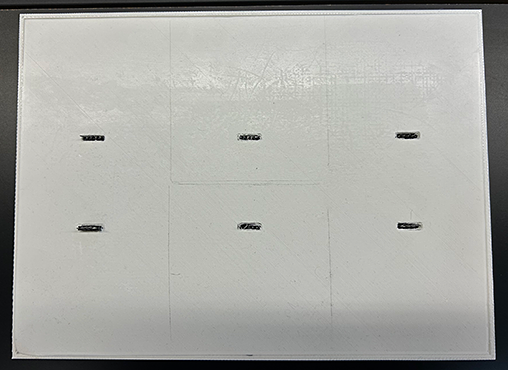}
         \vspace{-0.15cm}
         \subcaption[]{Core component top view}
         \label{fig:a}
     \end{minipage}
     \begin{minipage}{0.495\columnwidth}
         \centering
         \includegraphics[width=4.05cm]{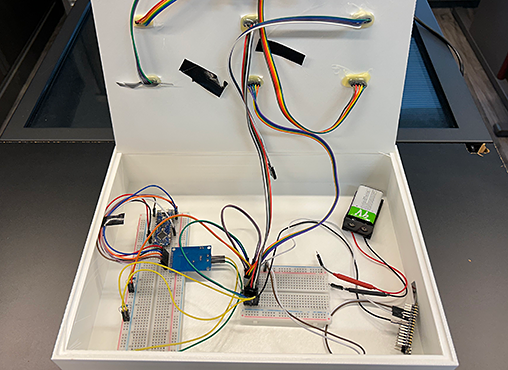}
         \vspace{-0.15cm}
         \subcaption[]{Core component internal view}
         \label{fig:b}
     \end{minipage}
     \begin{minipage}{0.495\columnwidth}
         \vspace{0.1cm}
         \centering
         \includegraphics[width=4.05cm]{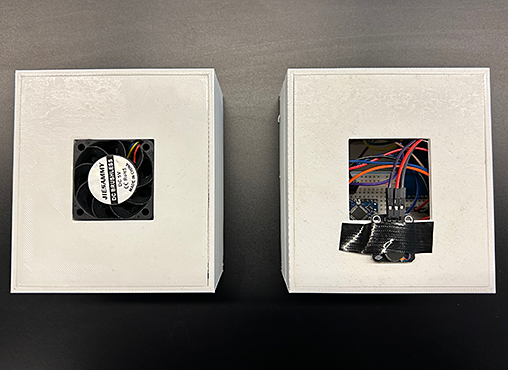}
         \vspace{-0.15cm}
         \subcaption[]{Module top view}
         \label{fig:c}
     \end{minipage}
     \begin{minipage}{0.495\columnwidth}
         \vspace{0.1cm}
         \centering
         \includegraphics[width=4.05cm]{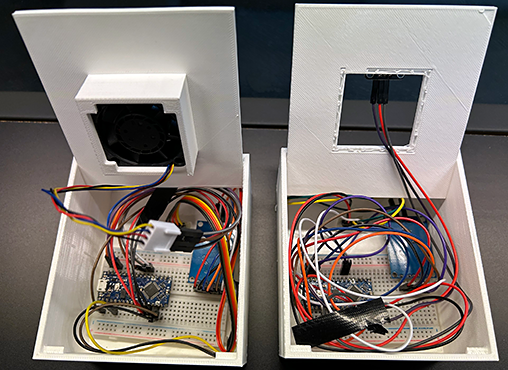}
         \vspace{-0.15cm}
         \subcaption[]{Module internal view}
         \label{fig:d}
     \end{minipage}
     \vspace{0.1cm}
     \begin{minipage}{0.99\columnwidth}
         \vspace{0.1cm}
         \centering
         \includegraphics[width=8.356cm]{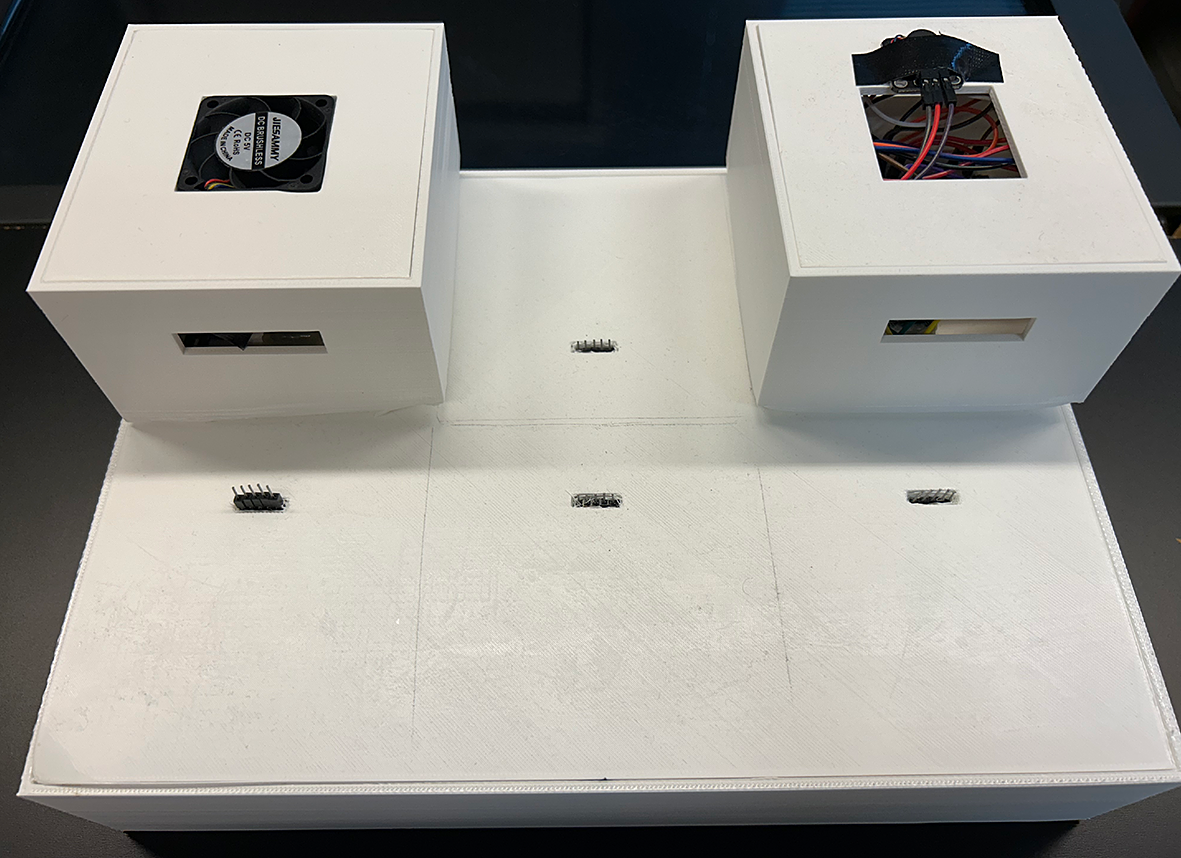}
         \vspace{-0.5cm}
         \subcaption[]{Core component with two connected modules}
         \label{fig:e}
     \end{minipage}
        \vspace{-0.25cm}
        \caption{Dynamic data physicalisation platform prototype}
        \label{fig:prototype}
\end{figure}

As illustrated in Figure~\ref{fig:e}, the core component of our prototpye consists of a 3D-printed box that has the size of exactly six modules arranged in a $2\times3$ grid. At the top of the core component, there are six connectors that a module can plug into. Inside the core component, one Arduino handles CAN communication while another Arduino is placed at each slot, as shown in Figure~\ref{fig:b}. Further, there is a battery providing power supply. Regarding the modules, we implemented two modules with different hardware representing physical variables: one fan and one vibration motor. Each module is a small box just fitting the required hardware, as highlighted in Figure~\ref{fig:c}. The rectangular holes on both sides of the module have two purposes. They provide airflow to prevent overheating and allow the fan to pull in air. This part of the module is meant to stay the same independently of the implemented physical variable. For the top part of the module, one can either use the provided parts or design some customised ones. 

A CAN message can transmit a maximum of 8~bytes. In our prototype, we encode the 8~bytes as follows: The first byte is used for the sender's address. The second byte indicates the type of message: \enquote{\code{n}} for a newly plugged-in module, \enquote{\code{h}} for a heartbeat message and \enquote{\code{s}} to set a specific physical variable to a new value. The remaining bytes in an \code{n}~message are \code{min} (describing the minimal value of a physical variable), \code{max} (describing its maximal value), \code{no} (specifying the number of values that can be encoded) and \code{index} (indicating the index number of a physical variable in the module when multiple physical variables are present). While the \code{h} message only has two bytes, the \code{s} message uses three more additional bytes, providing the address of the target module, the index number of the targeted physical variable in a module and the value it should be set to. 

When a module is connected to the core component, it first receives an address from the slot and is then ready to send \code{n} messages over the CAN bus. A module can send more than one \code{n} message if it implements multiple physical variables. The core component now knows which module is connected at which location and how to set its value accordingly. Similarly, the core component might send multiple \code{s} messages to set the values of multiple physical variables in the target module. Furthermore, the core component keeps track of whether a module gets disconnected by sending heartbeat messages.

As previously mentioned, we have not yet implemented the self-describing specification. Instead, we used a class instance to store the name, minimum and maximum values as well as the number of values that can be encoded by a physical variable. This structure perfectly fits the 8~byte requirement of a CAN message but is also flexible enough to add more attributes if needed in the future.

\section{Use Cases}
In the following, we describe two scenarios in which the proposed prototype could be used.

\subsection{Scenario 1: Researcher}
A researcher in the field of dynamic data physicalisation aims to experiment with how different physical variables impact end users' perception and comprehension of data. They use a dataset about global warming as the data input and prepare three different physical variables for the experiments: colour, temperature, and vibration. Our proposed prototype provides three different modules integrated with LEDs, heaters and vibrators, respectively.

For pre-setup, they turn on the platform (core component) and connect to a CSV file of global warming data. In the first experiment, they connect each LED module to the appropriate spot on the platform and then choose which data dimensions will be represented by which LED colours. After making their selections, the prototype renders a physical output accordingly. Finally, they ask users to explore the output and answer several questions about the dataset.

In the second and third experiment, they repeat the same steps but replace the LED modules with heater or vibrator modules. Thanks to our prototype's plug-and-play structure, they can easily disconnect unused modules and connect the desired ones. They can also combine different physical variables by connecting different modules to adjacent spots and choosing to use two physical variables to represent the same value.

\subsection{Scenario 2: End User}
A running enthusiast wants to make an artefact to display their exercise and health data. They use a sports watch to track their activities, with the data being uploaded to the cloud in real time. They turn on the proposed platform, connect it to their sports data and select two modules to represent two attributes: a shape-changing module for running distance and an audio module for heart rate. They connect the shape-changing module to the left side of the platform and the audio module to the right. They then run on a treadmill and observe how the customised artefact functions. As their running distance increases, the left module moves up, while the right module adjusts the music volume according to their heart rate. 

\section{Discussion and Future Work}
We proposed a plug-and-play architecture and platform that enables the rapid prototyping of dynamic data physicalisation applications. The combination of a CAN~bus and PLC component perfectly meets the requirements for the platform to be notified whenever a module is connected or disconnected, and ensures that all modules can communicate with the core component at any time. 

We have implemented a preliminary prototype of the proposed architecture. Our proof-of-concept prototype includes two different modules, providing vibration-based and airflow-based feedback. We also conducted a technical evaluation to verify the functionality of the prototype and the underlying communication protocol. As part of this evaluation, the two modules were plugged in and removed multiple times to ensure the core component correctly detected the connection and disconnection of modules. Further, we tested the core component's ability to set various values for physical variables on the modules and checked for correct behaviour when incorrect values were sent (e.g.~values outside the min/max range). All these tests were successful and performed as expected without any errors.

Although the presented prototype verified the feasibility of the proposed solution, there are still several improvements for future work. First, we used standard GPIO pins and cables as connectors, which were quite fragile and difficult to plug in. Magnetic connectors would be more convenient and user-friendly for developers or researchers working in the field, and we plan to investigate this option for the next version of the platform. Second, instead of using the PLC~component for recognising connected modules, we currently used a serial cable due to a lack of proper hardware. This issue might be resolved in a future iteration of the platform by developing our own PLC hardware for this functionality. Furthermore, a user interface for interacting with the platform is required. In our prototype, we used Arduino's Serial Monitor, a simple console for input and output. In a next step, a display showing the necessary information about the current status of the platform and allowing more efficient and user-friendly interaction and control is needed.

As discussed earlier, the presented modular and extensible hardware platform forms part of a broader vision of a framework for dynamic data physicalisation~\cite{signer2018}, where a data physicalisation grammar will inform the automatic transformation and mapping of data to physical variables, potentially implemented via the presented hardware platform. Finally, the self-describing specification for physical variables still needs to be generalised. This will allow researchers and developers to create their own modules and make them compatible with the platform by following the self-describing specification format for physical variables. While our presented platform has a regular matrix layout, in the future, the presented plug-and-play architecture might also be used with other customised layouts no longer following the regular layout of our initial prototype.
\section{Conclusion}
In this paper, we presented a solution to build a modular and extensible dynamic data physicalisation platform. The main contributions of this paper include a technical solution that encompasses the \emph{communication protocols and necessary hardware for a plug-and-play architecture}, making it easy and fast for other researchers and developers to construct dynamic data physicalisations. Further, the \emph{physical design of the platform} consists of a core component and modules that can be equipped with various sensors and actuators to implement specific physical variables, along with the necessary connectors for these modules. Based on our evaluation, we might investigate new magnetic connectors in the future. Additionally, we proposed the introduction of a \emph{self-describing specification} for physical variables, allowing other researchers to build their own modules to meet their needs and extend the platform as desired.

We also implemented an initial prototype to demonstrate the feasibility and usability. Although it still has some limitations that might be addressed in future iterations of the platform design, we are confident that the presented hardware platform can facilitate the rapid prototyping and development for researchers and developers in the field of dynamic data physicalisation. The proposed solution can be easily replicated due to the availability and low cost of the used hardware components. Note that in the future we plan to release our platform under an open source license and make it available to the research community. This will allow researchers and developers to benefit from the presented extensible hardware platform for dynamic data physicalisation by no longer having to start from scratch when realising a plug-and-play architecture and communication between components.

\balance

\bibliographystyle{ACM-Reference-Format}
\bibliography{TR-WISE-2025-01}

\end{document}